\documentstyle[proceedings]{crckapb} 

\def\la
  {\mathrel{\hbox{\raise.3ex\hbox{$<$}\llap{\lower.8ex\hbox{$\sim$}}}}}

\begin{opening}
\title{CARBON STARS AND
       NUCLEOSYNTHESIS IN GALAXIES}

\author{BENGT GUSTAFSSON~} 
\author{~NILS RYDE}
\institute{Uppsala Astronomical Observatory\\
           Box 515, S-751 20 Uppsala, Sweden}

\end{opening}

\runningauthor{B. GUSTAFSSON ~AND ~N. RYDE}

\begin{document}

\begin{abstract}
        The role of carbon stars in the build-up of chemical elements
in galaxies is discussed on the basis of stellar evolution calculations
and estimated stellar yields, 
abundance analyses of AGB stars, galactic-evolution models and 
abundance trends among solar-type disk stars. We conclude that the 
AGB stars in general, and carbon stars in particular, probably 
are main contributors of $s$-elements, that their contributions of flourine 
and carbon are quite significant, and that possibly their contributions
of lithium, $^{13}$C and $^{22}$Ne are of some importance. Also  
contributions of N, Na and Al are discussed. 
The major uncertainties that characterize almost
any statement concerning these issues are underlined.
\end{abstract}

\vspace*{-0.5cm}

\section{Introduction}

Any discussion of the role of carbon stars (C stars) in the chemical 
evolution  of galaxies raises 
a number of important questions such as: For which metal abundances and 
initial stellar masses do stars become C stars? How much mass does a star 
lose during its C star phase? What are the elemental compositions 
of these ejecta? 

C stars are brilliant and show easily recognizable spectral
characteristics. Thus, their occurrence and frequency in different galaxies
may be interpreted in terms of the properties (metallicity and age) of the 
stellar populations, if we know which stars become C stars.
Current answers to this question are, however, not very precise, 
although stars with masses somewhere 
in the interval 1.2 to 4\,M$_\odot$ may be a realistic answer. 
Studies of the distribution of C stars perpendicular to the Galactic
plane suggest typical masses in the interval 1.2 -- 1.6\,M$_\odot$ (Claussen 
et al. 1987; cf. also Groene\-wegen et al. 1995). For stellar masses 
greater than about 
4\,M$_\odot$ hot-bottom burning (HBB) is thought to prevent C star
formation (Boothroyd et al. 1993). 
In metal-poor populations, stars with even smaller masses 
may become sufficiently carbon enriched to show C star spectra,
while HBB may also be
effective in inhibiting C star formation at lower masses than for Pop.~I 
stars. An important issue in
this respect is the frequency of C-rich planetary nebulae (PNe) in the Galaxy,
which is so high (cf. Zuckerman \& Aller 1986; Rola \& Strasi\'nska 1994)
that a considerable fraction of all stars must become 
C stars in the end --- this is a strong reason for not increasing the 
lower limit of the mass range discussed here too much above 1\,M$_\odot$.

There is consensus that mass loss probably both sets the maximum luminosity
that a star 
achieves and determines the time it spends in its evolution up along the
AGB. The total mass returned to the interstellar medium 
is obviously the difference between the initial stellar mass
and the final remnant mass (typically about 0.65\,M$_\odot$ for the stars
discussed here). In order to determine the total mass lost from the
C stars we first need to know how much mass these stars have lost before
becoming C stars. Most authors assume that the total mass loss on the RGB
is on the order of 0.2\,M$_\odot$ for solar-mass stars, following arguments 
by Renzini \& Fusi Pecci (1988), while more massive stars --- being more 
compact and 
not undergoing the He core flash --- presumably lose 
less. Many studies also assume some continuous mass loss, specified by the
Reimers relation (with a suitable fitting parameter), followed at the end 
of the AGB by a ``superwind" with a steeper dependence of mass loss on 
stellar parameters.
Detailed models of pulsating stars (Bowen \& Willson 1991; 
Willson et al. 1996) demonstrate the inadequacy of simple parametrizations  
--- the mass loss rate turns out to be quite sensitive to stellar
parameters, including mass and composition, and this cannot be determined 
empirically from available observed mass loss rates because
the stellar parameters are not well known.

Since the atmospheres of 
AGB stars, and especially C stars,
are enriched in many elements by interior nucleosynthesis and various mixing  
mechanisms, 
they contribute substantial amounts of heavy elements to
the interstellar medium (ISM). 
(Also, the physical and chemical
properties of the ISM are affected by the grains produced by
C stars.) 
The main yield to the ISM from the C stars depends,
however, on 
the interplay between mass loss and dredge-up of processed material 
from the inner stellar layers. Neither of these processes is very well 
understood. In recent model
calculations dredge-up, as well as mass loss, is incorporated with free
parameters adjusted to fit some selected observational constraints, such as 
the luminosity distribution of C stars in the Large Magellanic Cloud (LMC). 
Major recent 
contributions of this type are due to Groenewegen \& de Jong (1993),
Groene\-wegen et al. (1995), Bl{\"o}cker
(1995) and Marigo et al. (1996). This may be a questionable procedure in 
view of the assumptions made concerning the dependence
of the adjustable parameters on stellar properties.

Key issues in the present discussion are when, 
during the AGB evolution with its 
gradual dredge-up of enriched material, the most significant mass loss 
occurs and how this depends on the metallicity of the
star. We are still far from definitive answers to these questions.
Here, a rather empirical discussion will be given,
based on the observed chemical compositions of stars and PNe.
For a detailed and more theoretical approach, based
on calculations of nuclear processing and mixing during the AGB with
adjustable parameters describing the dredge-up as well as with semiempirical  
recipes for mass loss rate as a function of luminosity and pulsation 
period, see Marigo et al. (1996). These authors also contribute tables of 
predicted yields, partially superseding those of Renzini \& Voli (1981). 
   
Another, more indirect but very significant basis for estimating
yields will also be used below: the trends in relative abundances 
of solar-type stars of different ages in the solar neighborhood. 
From such results (e.g. those of Edvardsson et
al. 1993) one can estimate relative time scales of the build-up of 
different elements, and thus separate the contribution
of elements from, say, slowly evolving low mass
stars as compared with contributions from rapidly evolving high mass 
stars through supernova (SN) explosions.

We shall here review the current knowledge, and lack of knowledge,
as regards the role of C star production of the 
nuclei of Li, C, N, O, F, Ne, Na, and Al as well as of 
the $s$\/-process elements. Our discussion will be based on
what is known from stars in the galactic disk and in the LMC, 
and it is important to stress
that the understanding concerning the mechanisms
of the formation, nucleosynthesis and mass loss of C stars is so
weak that extrapolations to stellar populations with other properties 
are highly
uncertain. In fact, the mixture of arguments, partly based on galactic, 
partly on LMC C stars, in contemporary discussions 
is in itself dangerous, but necessary.

\section{Nucleosynthesis}

\subsection{lithium}

As has been demonstrated by Reeves et al. (1990),
spallation of CNO nuclei in the interstellar medium,
although contributing $^{6}$Li, Be and B, cannot be the main process for
raising the $^{7}$Li abundance from the Pop. II plateau value found by 
Spite \& Spite (1982) to the more than 10 times higher value found in the
solar system and in young stars. The processes by which Li has been built
up are as yet not determined, but several processes have been proposed:
Li-production in AGB stars, 
the $\nu$-process in SNe of Type II (SNeII),
and production in novae. In a study based on models of
galactic chemical evolution and observed abundances of light elements,
Brown (1992) suggested intermediate-mass stars to be main contributors.  

Some C stars show very strong Li\,{\sc i}
resonance lines (McKellar 1940) and
these stars have been proposed to be a significant source of Li to 
the ISM (Scalo 1976). These rare stars, having 
logarithmic Li abundances, log $\epsilon$(Li), of about 10 times 
the solar system value or even greater
($\sim$\,4 on a scale with log $\epsilon$(H) set to 12),
are still not fully understood; attempts to 
explain them include
HBB and the Be transport mechanism of Cameron \&
Fowler (1971) (Sackmann \& Boothroyd 1992; Abia et al. 1993b).

Abia \& Isern (1996) found that the 
$^{13}$C abundances of galactic C stars correlate with their Li 
abundances; this may give a 
further clue to the origin of the super Li-rich phenomenon. These 
authors suggest that HBB might produce the abundances found if one 
invokes a more efficient convection than is usually assumed. This could 
decrease the lower mass limit of HBB to below 2\,M$_\odot$.
Alternatively, the plume mixing model of Scalo \& Ulrich (1973) might 
operate in these stars. 

The significance of C stars in general for galactic Li production is 
entirely dependent on the role of the super Li-rich 
stars; the more normal N stars have log $\epsilon$(Li)
ranging from --2 to 2 (Abia et al. 1993b) which is far too low
for the stars to be of significance 
in this respect. Key questions concerning the super Li-rich stars are: 
(1) How common are they? (Note that the conclusion of Abia et al.
that these stars may provide a significant fraction of 
the galactic Li is based on only one star!)
(2) How does their frequency depend on mass, 
metallicity and other parameters?
(3) What are their true Li abundances? (We note that the abundance 
estimates are severely dependent on model atmospheres, 
spectral synthesis and adopted C/O 
ratios for the models. Thus, Abia et al. (1993a) derive an abundance for
WZ Cas of log $\epsilon$(Li) = 5, while Boesgaard et al. (1996) find a 3
times lower abundance with the same grid of model atmospheres.)
(4) Are the super Li-rich stars overrepresented, or represented at all, 
among the dust-enshrouded stars with mass loss rates $>
10^{-5}$\,M$_\odot$/year?
(5) Do the super Li-rich stars represent relatively short episodes, 
followed by Li burning, and certain limited mass intervals --- 
as is suggested by the model calculations by
Sackmann \& Boothroyd (1992)
and Frost \& Lattanzio (1995) --- or could they possibly 
mark the ending stage of AGB evolution in general?

A conclusion in several studies is that the C stars 
probably do not contribute significant amounts of Li 
(cf. Matteucci et al. 1995), even though contributions 
as high as 30\% or more of the interstellar Li may be possible
if some super Li-rich stars also have high mass-loss rates (cf. Abia
et al. 1993b). 

An interesting issue is whether this situation is different for low 
metallicities --- i.e., in early phases of galactic evolution or in the 
present evolution of dwarf galaxies. A low metal abundance seems
to lead to HBB at smaller stellar masses (cf. Sackmann \&
Boothroyd 1992). 
On the other hand, the low metallicity stars may lose their mass
at lower rates, such that the Li produced may be burnt away before 
most of the mass loss occurs (Matteucci et al. 1995). 
Empirical support for this may be the 
Li abundance of log $\epsilon$(Li) $\approx 3$ derived for 
luminous AGB (S) stars in the 
SMC (with a metallicity 1/3 of solar) found by Plez et al.
(1993), which, in spite of being significantly higher than expected from
standard AGB evolution is still 
lower than values for the most Li-rich AGB stars in the
Galaxy.

In conclusion it seems probable that another process, 
such as the $\nu$-process in SNeII (Woosley et al. 1990), 
is the main mechanism responsible for Li formation. However, until we
understand the nature of the super-Li rich C stars better, more
definitive statements must be avoided. 

\subsection{carbon}

For some time, intermediate or low mass stars have been 
considered as probably important for the 
production of carbon in the Galaxy (see, e.g., Tinsley 1978). 
Sarmiento \& Peimbert (1985) estimated the carbon contribution from 
AGB stars to be on the order 
of 60--80\% of all galactic carbon. Their argument was essentially
based on the fact that models of SNe and novae suggested that
these objects could not produce more than a 
minor part of the carbon in the interstellar 
medium. Sackmann \& Boothroyd (1991) claimed, on the basis
of calculations of dredge-up of carbon in AGB model sequences,
that low and intermediate 
mass AGB stars are the dominant sources of carbon in the universe.
The more recent supernova models of Woosley \& Weaver (1995) and 
Thielemann et al. (1996) predict yields that are mutually different by 
about a
factor of two, the Thielemann et al. yields being lower due to a high 
rate for the $^{12}$C($\alpha$,$\gamma$)$^{16}$O reaction. 
Adopting the higher yields of Woosley 
\& Weaver (1995), Timmes et al. (1995) still find in their models of 
galactic evolution that SNe and novae cannot produce enough, so that the 
contributions from stars with a mass less than 11\,M$_\odot$ 
must be dominant. Timmes et al. also adopt 
the (now partly obsolete) yields from intermediate
and low mass stars of Renzini \& Voli (1981) and find that
the carbon abundance relative to iron, [C/Fe], was 
lower by about a factor of two than its present value in 
the early evolution of our galactic disk; then it increased above its   
present value and finally decreased again as a 
result of iron production in less massive SNeIa. 

This particular behavior of [C/Fe] relative to [Fe/H] is not seen
in the composition of solar-type disk stars of different
ages, according to Tomkin et al. (1995). Instead, these authors
find a steady decrease in [C/Fe] {\it vs.} [Fe/H], suggesting a less
dramatic variation of contributing sites during the history of the Galaxy.
In fact, the ratios of carbon relative to oxygen and $\alpha$-elements
(produced in SNeII) only vary slowly and smoothly 
with [Fe/H]. This suggests that the carbon may well have been produced 
in high-mass stars, 
although carbon does not follow oxygen in the halo (Tomkin et al. 1992),
nor in dwarf galaxies (Garnett et al. 1995).
The decrease of [C/Fe] with increasing [Fe/H] in the galactic disk
is very different from the variation of the $s$\/-element abundances 
with [Fe/H] (see also Edvardsson et al. 1993). 
This suggests that the $s$\/-elements
were formed in stars with considerably longer characteristic lifetimes, 
i.e. smaller masses.

Prantzos et al. (1994) studied the formation of oxygen and carbon in the 
Galaxy and found that, if the duration of the halo phase was on the 
order of 1--2 Gyears as is currently believed, 
intermediate or low mass stars should not have been the main 
carbon sources. Instead, these authors suggest that massive stars
contribute with metal-dependent yields (Maeder 1992). 
We note that the model calculations of Prantzos et al., with contributions 
added also from the intermediate and low mass stars 
according to Renzini \& Voli (1981), fit the results of Tomkin et al. (1995)
rather well, although the C/O ratios of comparatively metal-rich halo stars 
become too high. In that model, about 40\% of the 
carbon in the disk stars was contributed by intermediate and low mass stars.
Adoption of the yields of Marigo et al. (1996) would diminish
the discrepancy for the metal-rich halo stars but may produce too much 
carbon at late stages.
 
Other data suggest that the intermediate
and low mass stars may play a significant role in carbon
synthesis. Adopting the findings by Zuckerman \& Aller (1986) and Rola 
\& Strasi\'nska (1994) that about half of the PNe
have C/O $>$ 1, we find that these stars should contribute at least 
about 0.001\,M$_\odot$/year of carbon to the Galaxy.
(Here, we have used the PNe birth-rate of 
about 1 PN per year in the Galaxy estimated by Pottasch (1992), and a 
characteristic PN mass of 0.3\,M$_\odot$.) Assuming
the yields recently derived from a semi-empirical modelling 
of the AGB phase by Marigo et al. (1996) would presently give  
contributions on the order of 0.003\,M$_{\odot}$, 
depending on the star formation rate adopted. 
This contribution is of the same order of magnitude as 
the total present contribution by SNeII, assuming yields from Woosley 
\& Weaver (1995) and a SN rate of 3 per century.
In making this estimate we have assumed a closed box model; this 
may overestimate the 
significance of SNe relative to PNe, in view of the 
much higher expansion velocities
of material ejected in SNe. 
One should note that the number of C-rich PNe is so high
that they cannot only be produced by intermediate-mass stars; a significant
fraction of the C-rich PNe
must be formed by stars in the mass interval between 1 and
2\,M$_\odot$. 

If low-mass stars are assumed to produce significant amounts of carbon, 
further discrepancies between models of galactic chemical evolution and 
observed abundances 
may occur in the relative-abundance diagrams, such as those of Tomkin et al. 
(1995). 
Possibly, this conflict could be resolved by advocating more efficient 
carbon production by intermediate and low mass metal-poor stars, following
the discussion of Boothroyd \& Sackmann (1988) that suggests a more efficient 
dredge-up of carbon for models with low abundances of heavy elements. 
We note that the carbon yields by 
Marigo et al. (1996) are several times greater at a given
initial stellar mass for their metal-poor (Z=0.008) models than for their 
Pop. I (Z=0.02) models. (This is, however, dependent on the assumptions made 
in these models that the two adjustable third-dredge-up parameters, 
the minimum core mass for dredge-up and 
the dredge-up efficiency, are independent of metallicity, and that
the mass loss rate is dependent only on the metallicity through its effects 
on evolutionary tracks and pulsation period.)
Another possibility, which should be further investigated, is that the 
spectrum analysis of the high-excitation C\,{\sc i} lines in solar-type stars 
may give systematic errors, different for stars of different metallicities,
e.g. due to errors in the effective temperature scale, or errors due to 
inhomogeneities and departures from LTE (we note, however, the small non-LTE
effects found for carbon in the Sun by St\"urenburg \& Holweger 1990). 
The study of Andersson \& Edvardsson
(1994), using the low excitation [C\,{\sc i}] line at $8727$\,\AA, seems to 
verify the results of Tomkin et al. but the weakness of that line 
only admitted upper limits for most of the metal-poor disk stars in their 
sample. 
Further studies based on this line with higher S/N are underway. 

Do the C stars contribute significantly to galactic carbon? 
Number densities and mass-loss observations of bright, as well as 
dust-enshrouded, C stars in the Galaxy (Olofsson et al. 1993a, 1996;
Jura \& Kleinmann 1989) suggest a total contribution of about 
$2\times10^{-4}$\,M$_{\odot}$/year. The indication, from the 
statistics of PNe, that C stars may contribute one order of magnitude
more is the result of the fact that about every second PN is carbon rich. 
This illustrates the key significance, mentioned in the Introduction, of 
the question of the chemical composition of the stellar envelope just
prior to the extensive mass loss that ends the AGB evolution. 

We conclude that the significance of low and intermediate mass
stars for carbon synthesis is still unclear, and deserves further 
exploration. 

\subsection{the carbon isotope $^{13}$C}

Prantzos et al. (1996) have discussed the 
observations of carbon isotope ratios in the galactic disk and conclude that
$^{13}$C has probably a mixed origin, being produced both as a primary 
element --- presumably in intermediate-mass stars 
by HBB or by other processes where protons are 
mixed with $^{12}$C produced in the star by He burning --- and as a 
secondary element by CN burning, occurring in stars of all masses. 
The $^{12}$C/$^{13}$C ratios observed in 
C stars range in the interval 30--100,
with a pronouced peak around 50 (Lambert et al. 1986; a lower range, 15--40,
has, however, been obtained by Ohnaka \& Tsuji 1996, and these
differences need further exploration).  Excluded from this
is a minority of J-type stars with considerably lower 
ratios. Values around 50 are
consistent with the C/O ratios derived by Lambert et al.; i.e., the 
result may be explained as the consequence of 
mixing $^{12}$C to the surface layers if $^{13}$C is
left from the first dredge-up. These values agree reasonably well with 
those (30--40) derived for dusty C stars from mm-line observations by 
Kahane et al. (1992). These authors have
also observed the C-rich PN NGC 7027 and find a
$^{12}$C/$^{13}$C lower limit of 65. In fact, the difference
in C/O ratio between bright N stars and C-rich PNe (see Lambert et 
al. 1986) suggests that further $^{12}$C enrichment has taken place in the 
latter,
which would correspond to
an increase from 50 to about 80 in $^{12}$C/$^{13}$C. Although this agrees  
with the local ISM value, that may well be fortuitous --- 
little is known about the $^{12}$C/$^{13}$C ratios in PNe. One
should also note that the $^{13}$C ``produced" by normal
C stars is, as presumed above, just the result of the CNO burning and
the first dredge-up. That is, however, not necessarily the case; 
we note that Plez et al. (1993) find low 
$^{12}$C/$^{13}$C ratios for S stars in the Small Magellanic Cloud indicating
H-burning in the envelope in late evolutionary phases. 

One might ask what the role of the extremely $^{13}$C-rich J stars may be
in providing $^{13}$C. The evolutionary history of these stars is still not 
understood (cf. Lambert et al. 1986 for some suggestions). They constitute 
about 1/5 of the bright N-type stars in the magnitude-limited
Lambert et al. sample and have a $^{12}$C/$^{13}$C 
ratio of typically 5. They tend, however, to be systematically oxygen-poor;
therefore, the $^{13}$C abundances relative to hydrogen
are not more than typically
a factor of 5 higher 
than those of other C stars. The J-type stars do not show any tendency of
having higher present mass-loss rates than the rest of the  
N stars (Olofsson et al.
1993a). From this one would only conclude that the J stars at present
contribute approximately the same total amount of $^{13}$C as the rest of 
the visible N stars.
Similarily, if the J-type stars lose their envelopes without first
burning their $^{13}$C, and if the fraction of J stars among C stars of 20\% 
really is representative also of the population of immediate progenitors of 
C-rich PNe, which is highly questionable, 
the J stars could then deliver as much $^{13}$C as the 
sum of the rest of the C stars. There is at least one additional 
circumstance that
might suggest that the J-type stars play a significant role in this 
respect: a great fraction of 
the most luminous red AGB stars in the LMC are $^{13}$C-rich C stars 
(Richer et al. 1979), possibly indicating that a greater fraction 
than 20\% of the more massive C stars reach this 
stage close to the end of their AGB evolution, at least in metal-poor 
populations. However, there is also a population of low-luminosity J-type
stars present (cf. Richer 1981; Bessell et al. 1983) which does not
necessarily correspond to the latest phases of AGB evolution. 
Brewer (1996) recently discovered 7 J-type stars among 48 C stars in two 
M31 fields and found the J stars relatively faint. 

\subsection{nitrogen}

$^{14}$N is thought to be produced by equilibrium CNO-burning 
in the hydrogen burning shell, 
brought to the surface layers at the first
dredge-up and even more at the second dredge-up, occurring in stars of 
4--8 M$_{\odot}$ and expelled during later stellar evolution.
Timmes et al. (1995), however, conclude, on the basis of abundance trends 
among galactic stars and radial gradients of CNO abundances in H\,II 
regions in galaxies, 
that $^{14}$N has a strong primary component 
which they ascribe to low metallicity massive stars. Woosley \& Weaver (1995) 
show that it is possible to create this isotope in massive stars and claim 
that approximatively 25\% of the solar abundance can be ascribed to these 
sites. 
Vila-Costas \& Edmunds (1993) show evidence for a delayed primary and a
secondary component relative to oxygen, the delay indicating that 
intermediate mass stars are responsible. The low N/O ratio observed for a 
high redshift gas cloud by Pettini et al. (1995) also supports an 
intermediate-mass primary component. 
The secondary component is dominant at high metallicities. 
The bulk of the solar abundance of $^{14}$N is attributed to 
low and intermediate mass stars (see e.g. Timmes et al. 1995).

Empirically, the evidence that C stars
contribute a significant extra amount of N is weak.  
Lambert et al. (1986) find, from their analyses of CN lines
in N-type star spectra, that these stars are 
not very N-rich, and this result is strengthened by recent values of 
the dissociation energy of CN. Olofsson et al. (1993b) find an N abundance 
higher by about a factor of 5 or so from their analysis of HCN mm 
lines from the circumstellar envelopes of the same sample of N stars, but 
this may be due to an underestimated mass loss rate (cf. also Olofsson 
et al. 1996).
The C-rich  PNe are, according to Zuckerman \& Aller (1986),
generally not very enriched in N --- only very few in their list have 
N abundances
in excess of twice the solar value. Pasquali \& Perinotto (1993) list a 
mean nitrogen abundance of twice solar for their C-rich PNe of type
II and III, which represent disk stars of intermediate and low mass, 
while the mean
N abundance given for PNe of type I, that are more carbon-poor and 
represent younger, more massive objects, is three times greater. 
Kingsburgh \& Barlow (1994) give a mean nitrogen 
abundance of only 1.4 times solar for their 36 non-Type I PNe. 
Similar tendencies may be traced in the yields calculated by 
Marigo et al. (1996). It seems probable
that intermediate-mass stars that generally are prevented 
from becoming C stars by HBB may provide most of the 
nitrogen. We note that Brett (1991), in his analysis of 
luminous AGB stars in the SMC, finds strong CN bands, 
indicating C to N conversion by HBB.

\subsection{fluorine}

The galactic production of the single stable isotope 
of flourine, $^{19}$F, has been
a matter of discussion and is still rather unclear. 
Jorissen et al. (1992) 
identified lines of HF in the 
2 micron region for K, M, S and N giants and derived 
F abundances. 
They found that the F/O ratio in AGB stars increases with 
C/O, with
maximum values for stars with C/O ratios not too 
much in excess 
of 1 (SC stars). Their result seems to imply that the thermal 
pulses produce
fluorine, and they (cf. also Forestini et al. 1992) suggest as the most 
probable scenario that $^{13}$C($\alpha$,n)$^{16}$O reactions
produce neutrons, some of which are
captured by $^{14}$N to produce $^{14}$C and protons, which in turn are 
captured by
$^{18}$O: $^{18}$O(p,$\alpha$)$^{15}$N($\alpha$,$\gamma$)$^{19}$F. The 
flourine is then 
brought to the surface by convection following the thermal pulse. 
This scenario is
complicated by several circumstances, e.g. the new higher rate for 
the reaction $^{18}$O($\alpha$,$\gamma$)$^{22}$Ne which may deplete 
$^{18}$O enough 
to decrease the significance of 
the second part of the reaction chain (cf. Frost \& Lattanzio 1995). 
In any case, the over-abundances of F found in SC and N stars by Jorissen 
et al.
(1992), ranging from 3 to 30 times the solar values, indicate that these
stars may well be major contributors to the galactic flourine; 
an alternative 
is production in SNII (Woosley et al. 1990; Timmes et al. 1995). 

An interesting way of 
clarifying this issue further may be to determine the flourine 
abundances in
less evolved M stars with different metal abundances, 
since different scalings of, say, [F/O] with [Fe/H] may be
expected for the different production sites.

\subsection{neon}

The dominant neon isotope in the solar system, $^{20}$Ne, is probably 
mainly produced by SNeII (cf. Woosley \& Weaver 1995; Timmes et al. 
1995). The SNe yields of $^{22}$Ne are about 
one order of magnitude lower than those of $^{20}$Ne which is
consistent with the isotopic ratios observed in the solar system.
However,  $^{22}$Ne may be
formed at thermal pulses in AGB stars
in the convective intershell through $\alpha$-capture 
on $^{14}$N, which after $\beta$-decay leads to 
$^{18}$O($\alpha$,$\gamma$)$^{22}$Ne
(Boothroyd \& Sackman 1988; Gallino et al. 1990). In the AGB model 
calculations of Marigo et al. (1996), considerable amounts of $^{22}$Ne 
are produced and ejected, in particular for models with masses around 
2.5\,M$_{\odot}$ which get most
carbon rich. The predicted $^{22}$Ne yields are fairly independent 
of metallicity (though
higher for Z=0.008 than for Z=0.02), and so high that the AGB stars 
are suggested to 
be a major contributor of galactic $^{22}$Ne. In particular, the 
envelopes of C stars are predicted to have a $^{20}$Ne/$^{22}$Ne ratio
considerably less than 1, even close to 0.1. 

The Ne enrichment expected in C-rich PNe according to these results 
was not traced in the sample of southern PNe studied by Kingsburgh \& Barlow 
(1994), nor can it be seen in the sample of PNe from the Magellanic Clouds 
studied by Leisy \& Dennefeld (1996). However, Corrardi \& Schwarz (1995) 
found a sample of bipolar PNe to be enriched in Ne and Marigo et al. (1996)
suggest that the build-up of $^{22}$Ne could be the reason for this.
We note that Lewis et al. (1990) and Nichols
et al. (1993) find consistently low isotopic ratios in gas-rich
meteoritic SiC grains. 

\subsection{sodium and aluminum}

Na and Al are, as well as Mg, generally identified as products 
of Ne and C burning in massive
stars while the production in SNeIa is probably small (Nomoto et al. 1992). 
The synthesis of Na and Al is controlled by
the neutron flux during Ne and C 
burning which in turn is dependent on the initial metallicity and primarily
on the initial O abundance ($^{16}$O being converted to $^{22}$Ne, 
via $^{14}$N,  
in He-burning and the extra neutrons in $^{22}$Ne, 
when being liberated, are essential
to the formation of Na and Al). Therefore, one expects a rapid increase of 
Na/Mg ratios with Fe/H, and such a tendency is also apparent in the 
calculations of Timmes et al. (1995).  
For the most metal-rich 
solar-type stars the positive correlation between Na/Mg and Fe/H
is well established (Feltzing \& Gustafsson 1996). 
However, the observations of disk stars by Edvardsson
et al. (1993), as well as observations of halo stars, do not show this 
tendency very clearly (Timmes et al. 1995).  
Edvardsson et al. find a tendency for the abundance ratio of 
Na/Mg to be smaller in the inner Galaxy than in the outer, at a given 
Mg/H, for disk-stars more metal-poor than the Sun.
This suggests that Na and Mg, respectively, were 
formed in stars of different types. 
Thus, there might be  
additional sources of sodium and possibly of aluminum. 

There is some additional evidence that intermediate-mass stars or even
low-mass stars might contribute in these respects. In intermediate-mass
stars sodium may be synthesized in the hydrogen-burning shell through the
neon-sodium cycle, and by proton captures on $^{22}$Ne (Denisenkov 1989; 
Denisenkov \& Denisenkova 1990; Langer et al. 1993). On longer
time-scales, i.e. smaller masses, proton capture on $^{20}$Ne may also be
significant, and the results may be brought to the surface if efficient 
mixing is at hand. Observationally, yellow field supergiants
appear sodium rich (Boyarchuk \& Lyubimkov 1985) but
they hardly contribute significantly to the galactic Na.
A fraction of low mass red giants 
in globular clusters show enhanced  Na and Al and are correspondingly
poor in O (Norris \& Da Costa 1995 and references given therein).
Recently Cavallo et al. (1996) have attempted to model this, by following 
the nucleosynthesis in low-mass red giant model sequences, and they find 
substantial Na-enrichments in the region above the hydrogen-burning 
shell throughout the giant branch, independent of metallicity. For the 
lowest metallicities Al is also produced. With suitable mixing processes 
these elements will be visible on the surface and may even, after mass 
loss, contribute
to the global enrichment of the Galaxy.

There is, however, no strong argument for this low-mass star production
of Na and Al to be linked to the existence of
C stars. Conversely, the meager abundance analyses that exist do not
support such a hypothesis; e.g. Lambert et al. (1986) find normal Na/Ca 
ratios on the basis of 1 line for each element in the 2 micron region. 

There is an interesting possibility that AGB stars
contribute radioactive $^{26}$Al (Kudryashov \& Tutukov 1988), 
decaying to $^{26}$Mg with emission of 1809 keV photons, detected in space.
Guelin et al. (1995) have recently found some evidence for $^{26}$AlF 
millimetre
emission from the carbon-rich envelope of IRC+10216. 
Huss \& Wasserburg (1996) comment on the finding of high $^{26}$Al/$^{27}$Al 
ratios in some SiC grains. The high temperatures needed to produce $^{26}$Al 
suggest rather high mass stars as sites if HBB is invoked. This is,  however, 
not easily reconciled with other isotopic ratios.

\subsection{the \lowercase{{\it s}}\/-elements}

The early discovery by Merrill (1952) of the unstable element technetium 
in S stars was a clear indication that these stars could be
a source of the $s$\/-process elements in general (see also Cameron 1955).
AGB stars are now generally found to be more or less $s$\/-element rich.
C stars are believed to be the most enriched (see below). 

Lambert (1992) gives several arguments in favor of the idea that  
$s$\/-elements are produced in AGB stars of low mass 
($\la 3$\,M$_\odot$) rather  
than in stars of intermediate mass. In intermediate-mass AGB stars, 
the temperature 
in the He-burning shell is high enough to ignite the 
$^{22}$Ne($\alpha$,n)$^{25}$Mg neutron source. 
That would lead to an excess abundance of
$^{25}$Mg (cf. Lambert 1991, and references therein). 
Smith \& Lambert (1986) found no $^{25}$Mg enhancement in their seven
randomly selected $s$\/-element enriched stars, which suggests that 
these were all low-mass AGB stars. 
In these stars the $^{13}$C($\alpha$,n)$^{16}$O reaction is the probable
neutron source.

The neutron density at the $s$\/-process site can be determined 
from observations of isotope abundances at certain branching points
of the $s$\/-path in the chart of nuclides. Two useful branching points 
are the ones at $^{95}$Zr and $^{85}$Kr. Lambert et al. (1995)
found no evidence of $^{96}$Zr in ZrO bands in S star spectra, 
implying a neutron density of less than $5\times10^8$ cm$^{-3}$,
much less than what is expected in regions where
the $^{22}$Ne neutron source operates. This again suggests that 
intermediate-mass AGB stars are not
the major contributors of $s$\/-processed elements in our Galaxy. 

Until recently, it was believed that the  $^{13}$C neutron source was
in operation during the thermal pulses of 
low-mass stars. 
Straniero et al. (1995) have, however, found from an 
evolutionary model sequence (for 3\,M$_\odot$) that the $^{13}$C 
neutron source is activated during the 
intervals  {\it between} the thermal pulses. This suggests 
a lower neutron density 
($10^{7}$ cm$^{-3}$), and that the $^{13}$C is consumed during the 
interpulse period. Semiconvection is essential for the $s$\/-process 
(see Sackmann \& Boothroyd 1991, and references therein); this will
allow mixing of protons into the carbon pocket of the thermal pulses 
leading to fresh $^{13}$C. Predictions from the models of Straniero et 
al. regarding one of the signatures of the low neutron density, the
Rb/Sr ratio which depends on the $s$\/-path at the $^{85}$Kr branch, 
are in agreement with recent observations 
of MS and S stars by Lambert et al. (1995). 

Mean neutron exposures deduced from observations of 34 MS and 
S stars by Smith et al. (1987) suggest that the solar system 
$s$\/-element distribution could arise from a mixture of MS, S and 
C stars which return 
$s$\/-element enriched material to the ISM. Their $s$\/-element excesses 
are for MS stars a factor of 2--3, for S stars 4--6 and for C stars 5--10. 
Note that the latter abundances are based on the 
uncertain abundance analysis of the crowded visual spectra by 
Utsumi (1985). 
These figures, in combination with estimated total mass loss, are high 
enough to explain the $s$\/-element abundance in the solar system.
Also, Parthasarathy (1996) found $s$\/-element enhancement relative to the 
Sun of 2--40 for some post-AGB stars.

There is also other more indirect, though strong, empirical evidence 
for the $s$\/-elements being formed mainly in low-mass AGB stars. Edvardsson
et al. (1993) showed, from studies of the 
composition of solar-type disk stars of different ages, that the major 
contributors to the enrichment of $s$\/-elements in the disk are 
stars with characteristic evolutionary times $>3 \times$ 10$^9$ years,  
i.e. low-mass stars. This is significantly longer than the time scale 
of the iron
production from SNeIa.  The low-mass contribution to the $s$\/-elements in 
the galactic disk is also verified by Pagel \& 
Tautvai\v{s}ien$\dot{\rm e}$
(1996). Moreover, from abundances in the halo they also trace a 
contribution of 
unknown origin but with a time scale characteristic of stars of 
$\sim8$\,M$_\odot$.

The third dredge-up may be even more significant for populations 
with lower metallicity than solar (cf. Wood 1981) and provide strong 
$s$-element enrichment. We note that Kipper et al. (1996) found three 
presumably intrinsic C stars of Pop. II to have 
$s$\/-element enhancements of 1--3 dex.

We conclude that mass loss from precursors of 
C-rich  PNe are most probably a main 
source of the $s$\/-elements in galaxies.

\section{Conclusions}

It seems most probable that the AGB stars in general, and C stars in 
particular, are the main contributors of $s$\/-elements; probable that 
their contributions of flourine and carbon are  
quite significant; and possible that 
their contributions of lithium (from the super Li-rich C stars), of 
$^{13}$C (from J stars) and of Ne (i.e. $^{22}$Ne)
are of some importance. However, all of these statements
are more or less uncertain, and the uncertainty is also great
concerning 
the role of intermediate and low mass stars in the production of 
N, Na and Al. This is because a number of relatively fundamental 
questions concerning
C stars still remain unanswered. In order to improve the situation 
new and more detailed and reliable abundance
analyses for C stars and  PNe are needed (also in nearby 
galaxies with different metallicities), as well as 
better models of nucleosynthesis and dredge-up in AGB stars. 
We also need to know more about mass loss --- how much matter the stars 
eject, as a function of initial mass and composition, 
as well as when the stars lose mass relative to the time when 
their envelope enrichment occurs. 
Further studies of abundances in less evolved stars of different
stellar populations are also of great significance. As regards the 
nucleosynthesis of two of the most important elements, 
carbon and nitrogen, more work on
stars more massive than normal C stars
seems also important.   

The situation reminds us of one of the stories about the famous Nasrettin 
Hoca 
of Konya, a place not very far from Antalya. Nasrettin saw ducks on the 
lake shore and, thinking of his dinner, tried to catch one but it jumped 
into the lake and swam out of reach. The next one did the same. After 
several attempts he 
took a spoon and sat down at the shore and started eating water. 
Some friends passed and asked him what he was doing. He answered: ``I am 
having duck soup."

We have tried to ``catch" the evasive secrets of the C stars with different,
more or less sophisticated methods. However, in spite of much progress
presented at this meeting, the solutions of the problems seem to be at some
distance. This partly reflects the experience, so common in science,
that phenomena tend to become more complex
when studied. In any case, more systematic investigations
are now needed. If we undertake
them, it seems that we shall get the most important answers long before 
Nasrettin Hoca has been able to walk to the center of his lake and catch 
his ducks. 

\vspace*{0.2cm}
John Lattanzio,  
Bernard Pagel, Lee-Anne Willson and Bob Wing
are thanked for valuable suggestions, and Martin Asplund and Kjell Eriksson 
for comments on the manuscript.

\newcommand{\MN}{{\it Mon. Not. Roy. Astron. Soc.}}
\newcommand{\Aa}{{\it Astron. Astrophys.}}
\newcommand{\Aas}{{\it Astron. Astrophys. Suppl. }}
\newcommand{\AaR}{{\it Astron. Astrophys. Rev.}}
\newcommand{\Apj}{{\it Ap.\,J.}}
\newcommand{\Apjs}{{\it Ap.J. Suppl.}}
\newcommand{\Aj}{{\it A.\,J.}}
\newcommand{\PASP}{{\it Publ. Astron. Soc. Pacific}}
\newcommand{\Araa}{{\it Ann. Rev. Astron. Astrophys.}}

\end{document}